# Effects of the grain size and shape on the flow stress: a dislocation dynamics study

M. Jiang[a,b], B. Devincre[a], G. Monnet[b]

[a] LEM, CNRS-ONERA, 29 Avenue de la Division Leclerc, BP 72, 92322 Chatillon, France

[b] EDF-R&D, MMC, Avenue des Renardières, 77818 Moret-sur-Loing, France

Abstract

Dislocation dynamics simulation is used to investigate the effect of grain size and grain shape on the flow stress in model copper grains. We consider grains of 1.25 – 10 µm size, three orientations (<135>, <100> and <111>) and three shapes (cube, plate and needles). Two types of periodic aggregates with one or four grains are simulated to investigate different dislocation flux at grain boundaries. It is shown that in all cases the flow stress varies linearly with the inverse of the square root of the grain size, with a proportionality factor varying strongly with the grain orientation and shape. Simulation results are discussed in the light of other simulation results and experimental observations. Finally, a simple model is proposed to account for the grain shape influence on the grain size effect.

Keyword: dislocation dynamics, Hall-Petch, grain size, grain shape, plastic deformation

## 1. Introduction

Intensive research has been dedicated to the effect of grain size on the mechanical properties of metals and alloys (Armstrong, 2014; Cordero et al., 2016; Lasalmonie and Strudel, 1986; Li et al., 2016). At the beginning of 1950, Hall (Hall, 1951) used the dislocation pileup model proposed by Cottrell (Cottrell, 1953) to explain the relation observed experimentally between the lower yield stress of mild steels and the grain size, *d*. This relation is recognized today as the Hall-Petch (HP) law:

$$\sigma_y = \sigma_0 + Kd^{-1/2}, \qquad (1)$$

where $\sigma_0$ is a reference stress generally called friction stress and *K* the HP constant. In the following, the quantity $Kd^{-1/2}$ is called the HP term. Some years later, Petch (Petch, 1953) has shown that the cleavage stress of mild steels also varies consistently with the HP law and Armstrong *et al.* (Armstrong et al., 1962) provided evidence that the HP law applies to the whole stress-strain curve and remains valid for most polycrystalline metals and alloys. It was also shown by Conrad and Schoeck (Conrad and Schoeck, 1960) and Petch (Petch, 1958) that the HP law applies to BCC metals at low temperature, with a temperature quasi-independent HP constant. In some experiments, *K* is found to be constant or to slightly decrease with deformation in BCC (P. Pechkovskii et al., 1989; Tsuchida et al., 2008), FCC (Feaugas and Haddou, 2003; Meyers et al., 1995) and HCP (Chia et al., 2005) metals and alloys. However,

Cordero *et al.* (Cordero et al., 2016) have shown that some pure metals show an opposite behavior with increasing K with deformation. On the other hand, Li *et al.* (Li et al., 2016) have shown that statistical analysis of the ensemble of experimental data on the Hall-Petch effect does not allow discriminating between the dependencies on $1/\sqrt{d}$, $1/d$, $\ln(d)/d$, etc. This has been attributed by Li *et al.* to the dispersion of results and the experimental errors in determining the grain size.

While the grain size effect applies to many mechanical properties of crystalline materials (including microhardness (Hughes et al., 1986)), several phenomenological rationalizations are possible (Hirth, 1972). Among the different interpretations, one can cite *dislocation pileup* models (ex. (Navarro and de los Rios, 1988; Smith and Worthington, 1964)), which refer to the original mechanism proposed by Hall and assume a stress concentration at the head of the dislocation pileup leading to the deformation propagation to the neighbouring grain by either activating dislocation sources in the next grain or enabling the transmission of dislocation across the grain boundaries (GBs). The *incompatibility-type* model relies on the original work of Ashby (Ashby, 1970) who introduced the concept of Geometrically Necessary Dislocations (GNDs) versus Statistically Stored Dislocations (SSDs). Such models interpret the HP effect as a result of the GND accumulation, necessary to accommodate strain incompatibility between grains. In this approach, the HP constant *K* is proportional to the square root of the plastic strain. However, Cordero *et al.* (Cordero et al., 2016) have shown that even in the cases where *K* increases with deformation, the increasing rate is not found to follow a square root. Also, Li *et al.* (Li et al., 2016) has pointed out another serious concern about this type of models: as K goes to zero in the absence of plastic deformation, the grain size effect must vanish at the threshold of plastic strain. Such behavior is not supported by experimental observations. A third type of models assumes a composite behavior of polycrystals where the grain interior and a layer bounding GBs have different strength (Kocks, 1970). Plastic flow is initiated in the GBs region owing to the incompatibility in elastic deformation of adjacent grains (Benson et al., 2001; Meyers and Ashworth, 1982). In another approach, Li (Li, 1963) (see also (Bata and Pereloma, 2004)) suggested that grain size effects are induced by dislocation emission from GB ledges. More recently, Sinclair *et al.* (Sinclair et al., 2006) proposed a new model accounting for dislocation storage at GBs and for their interaction with dislocations on the opposite side of the grain. The resulting elastic screening decreases the back stress, which in turn predicts a decrease of *K* with deformation.

A strong limitation of all existing models is that they do not provide a detailed description of dislocation arrangement and interactions in the vicinity of GBs. Numerical simulations are then needed to develop more physically justified models. Molecular dynamics are frequently used to analyze the GB response to one or few impinging dislocations (e.g. (Dewald and Curtin, 2007; Wang et al., 2014)). But the collective dislocation properties responsible for the HP effect can only be investigated with larger scale simulations such as discrete Dislocation Dynamics (DD). Based on 2D-DD simulations, Biner and Morris (Morris, 2002) studied the evolution of the flow stress in polycrystals with one slip system per grain and grain size ranging from 2–16 μm. Lefebvre *et al.* (Lefebvre et al., 2007) made simulations with grains in the range 0.5–2 μm size and two slip systems per grain. In both studies, the yield stress was found to vary as $d^n$, with an exponent *n* rather different from -½. 3D-DD simulations appear to be the only approach for quantitative investigations. Very few 3D-DD simulations of polycrystalline aggregate have been reported. Ohashi *et al.* (Ohashi et al., 2007) studied the effect of grain size on the stress necessary to activate Frank-Read sources. Their results suggest that sources with lengths close to one third of the grain size are the first to be activated. This effect was accounted for by introducing an addition hardening term proportional to the inverse of the grain size. On the other hand, the accumulation of GNDs at GBs was reproduced by Zhang *et al.* (Zhang et al., 2014) and connected to strain gradient in crystal plasticity. Only, De Sansal *et al.* (de Sansal et al., 2009) investigated the plasticity of a periodic polycrystalline aggregate made with regular polyhedral grains. The calculated yield stress in these simulations was found to vary with a HP exponent *n* between -½ and -1. In addition, this study underlined the absence of dislocation pileups and intensive cross-slip activity in the simulations with grain sizes lower than 1 μm. On the other hand, using 3d-DD simulations, El-Awady (El-Awady, 2015) has shown that, depending on the initial dislocation density, size effects may result from dislocation starvation, single-source strengthening or exhaustion hardening.

Remarkably, most published studies are concerned with the HP effects in polycrystals made of homogeneous equiaxed grains. Nevertheless, many materials of technological importance are composed of heterogeneous and non-equiaxed grains, such as bainitic and martensitic steels. Using the self-consistent scheme, Berbenni et al (Berbenni et al 2007) have shown that, not only the average grain size, but the type of the grain size distribution function and its standard deviation strongly affect the flow stress as well. The heterogeneity in grain size basically reduces the sensitivity to grain size. On the other hand, although grain shape is suspected to have strong effect on the flow stress (Van Houtte, 1982), little is known about the influence of grain shape on the HP Law (Delannay and Barnett, 2012; Hansen et al., 2010). Using DD simulations, only Yellakara and Wang (Yellakara and Wang, 2014) investigated the response of polycrystalline thin films and reported a variation of the HP exponent with the simulated volume shape. In the latter study, only cubic and hexagonal shapes were investigated, with close aspect ratios.

Giving the above underlined difficulties in measuring, interpreting, simulating and predicting the grain size effects in polycrystals, it is necessary to split the problem into different issues: (i) the effect of grain size and shape on the flow stress of individual grains and small aggregates with impenetrable boundaries as a function of the loading direction, (ii) the effect of dislocation – GB interactions such as absorption, repulsion, transmission, etc. (ex. (Lee et al., 1990; Shen et al., 1988)) and (iii) the elasto-plastic interactions between adjacent grains (ex. (Raabe et al., 2003; Sachtleber et al., 2002)). Clearly, the Hall-Petch effect cannot be correctly addressed without accounting for these three issues together. However, these features cannot be investigated by any individual simulation technique because they pertain to different scales. Feature (ii) involves the GB atomic structure and disorientation, while feature (iii) requires simulations at the macroscopic scale of a representative volume element. In contrast, feature (i) is related to the collective dislocation behavior inside one grain, which is the scope of mesoscopic simulations such as dislocation dynamics. The complete investigation of the Hall-Petch effect is thus still beyond the reach of one numerical simulation method. Many simulations at the three relevant scales are still necessary to provide a comprehensive picture of grain size effect.

In this paper, we report on 3D-DD simulations at the mesoscopic scale, pertaining to feature (i). It must be noted that the assumption of impenetrable GBs used in the present study does not give a fully description of the dislocation-GB interactions. Such proposition is made for reason of simplicity and because the development of more realistic constitutive rules accounting for all the phenomena observed at GBs in experiments or in atomistic simulations does not exist today. The development of such rules goes far beyond the goal of the present study. Still, 2D and 3D-DD simulations, accounting for dislocation transmission at GBs, have been proposed with some success (Zhou and Lesar, 2012; Quek et al., 2014; Fan et al., 2015; Burbery et al., 2017) to model plastic strain hardening in ultrafine-grains polycrystalline aggregates. These studies are considering grain size smaller than 1.5μm and therefore dislocation dynamics at very high stress. Contrarily, the present investigation considers grain sizes from 1.25 to 10μm and small plastic strain up to 0.2%. Hence, the stress on the dislocation arrested at GBs is always relatively low. This is why, the assumption of impenetrable GBs is expected to be an appropriate solution to investigate the Hall-Petch effect. The objective is to shed light on the evolution of the flow stress of one grain and small periodic clusters of grains as a function of the grain/cluster size, shape and orientation. In particular the collective dislocation properties are explored and a simple method is proposed to account for the grain shape effect. The paper is organized as follows: simulation technique and conditions are given in Section 2; then DD calculations on the grain size effect (Section 3) and on the grain shape effects (Section 4) are presented. The obtained results are discussed in Section 5 and general conclusions are summarized in section 6.

## 2. Simulation technique and conditions

The 3D-DD simulation code used in this study is *microMegas*. The basic features of this code are presented in reference (Devincre et al., 2011). In the following, only features important for the present simulations are specified. Pure copper is the model FCC material considered in the study. As illustrated in Figure 1a, initial dislocation microstructures are built up from random distributions of dipolar loops with square shape inside the simulated volumes. Every loop is composed of two pairs of edge segments belonging to a given slip system *s* and two edge segments belonging to the collinear system of *s*, i.e. sharing the same Burgers vector in a different glide plane. Dipolar loop distribution is preferred here to a distribution of Frank-Read sources, because it avoids infinitely strong dislocation pinning points that come with the use of sources (Mohles, 2001) and favors the emergence of a realistic 3D dislocation networks. At the beginning of simulations, the initial dislocation configuration is relaxed to generate an energy-minimized microstructure where dislocation segments are free to form junctions or annihilate. This relaxed 3D interconnected microstructure is thus closer to the "real" microstructure of un-deformed crystals and avoid using a distribution of Frank-Read sources with arbitrary length. The faces bounding the simulated volume are perpendicular to the <100> axes of the FCC lattice. Tensile tests are usually simulated by imposing a constant uniaxial strain rate in three specific crystalline directions ([100], [111] and [135] directions) in order to investigate different slip dynamics in single and multi-slip conditions. Unless specified, all values of the yield stress $\sigma_y$ are taken at 0.1% of plastic strain. The copper elastic constants, the Burgers vector and the Schmid factors calculated for the 3 tensile directions are listed in Table 1.

| Shear modulus $G$ (GPa) | Poisson ratio $\nu$ | Burgers vector $b$ (A°) | Lattice friction $\tau_f$ (MPa) | Schmid factor $m_{001}$ | Schmid factor $m_{111}$ | Schmid factor $m_{135}$ |
|---|---|---|---|---|---|---|
| 42 | 0.431 | 2.5525 | 2.5 | 0.408 | 0.272 | 0.49 |

Table 1: Main 3D-DD simulation parameters used to model Cu plastic strain

To make sure that simulations are close to quasi-static conditions, all computations are made at relatively low strain rate amplitude. It must be noted here that the strain rate used in simulations must be decreased when increasing the grain size since a fixed dislocation displacement produces less plastic deformation. The strain rate solutions used with each grain size are given in Table 2. Tests have been made systematically to control that the simulation results are not modified when the simulated strain rate is decreased by a factor 2.

Screw dislocation segments are allowed to cross-slip following a framework described in (Raabe et al., 2003). As previously discussed in (Dewald and Curtin, 2007), cross-slip is a key mechanism in the simulation of polycrystals with grains in the micrometer range. This mechanism is needed to avoid rapid exhaustion of dislocation sources when artificial pinning points are not considered in the initial dislocation microstructure.

## 2.1 Boundary conditions

Periodic conditions (PC) are applied to the simulated volume in all calculations and two types of boundaries are used. With the *transparent boundary* (TB) condition, dislocations are free to cross the simulated volume borders and they automatically reinter the simulated volume from the opposite face due to the PC. Alternatively, the *impenetrable boundary* (IB) condition confines dislocation motion inside the grain. Dislocations inside these grains are then immobilized when they reach an IB. Hence, TB condition is used to investigate the properties of an infinite *single crystal* (or bulk crystal) and the IB condition is employed to simulate plastic slip in grains with non-penetrable boundaries.

In this study two sets of calculations that make use of IB condition are performed. The first geometry is made with *one grain* occupying the total simulated volume. This simulation condition reproduces a polycrystal made of one periodic grain, which is named as the "*one-grain polycrystal*" simulation. Comparison between the *one-grain* polycrystal simulations and *single crystal* simulations that make use of TBs is illustrated in Figure 1. In the *one-grain* simulations, by virtue of the PCs, dislocations accumulating against the IBs experience elastic interactions with dislocations accumulated on the other side of the IB. Obviously, this feature is a strong limitation of the *one-grain* simulations since in real polycrystals, dislocations on one side of a GB experience interactions with dislocation belonging to an adjacent grain with different orientation and therefore different slip activity.

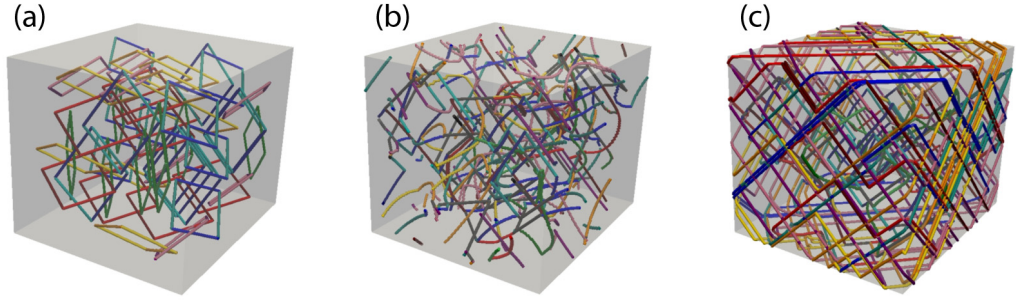

Figure 1(a): example of an initial dislocation configuration with a random distribution of dipolar loops; each color corresponds to a specific slip system. (b) and (c) are the dislocation microstructures obtained after deformation using transparent boundaries (TBs) and impenetrable boundaries (IBs), respectively.

In order to check the pertinence of the *one-grain* simulations, a second type of geometry with larger simulation volumes made of four grains is also considered (see Figure 2). In these "*four-grain* simulations", adjacent grains are loaded along different axes and contain initially different distributions of dislocation loops. This results in different slip activity on both sides of the IBs.

## 2.2   Grain size and initial dislocation density

To ideally simulate the HP effect, the influence of grain size on the flow stress should be computed, keeping all other microstructure features (such as dislocation density) unchanged. It turns out that this requirement is hard to fulfill for the following reason. The initial dislocation density is given by $\rho = (n \times 4l)/d^3$, where $n$ is the number of dislocation loops with a squared shape randomly inserted in a grain; $l$ the dislocation loop size and $d$ the grain size (see Fig. 1a). As the grain size is increased by a factor of 10 in the simulations, considering initial loops size independent of the grain size imposes a variation up to a factor of 1000 in the dislocation number between simulations to keep the dislocation density unchanged. Hence, working at constant dislocation density leads to strong computational time issues as well as geometrical difficulties to set up a realistic 3D dislocation network whatever the grains size. Rather, it is simpler to fix an initial loops size proportional with the grain size and to keep their number n and distribution unchanged. This simpler solution, similar to the one previously used in micro-pillars simulations (see for instance El-Awady, 2015), allows for simulations at different grain size with a dislocation density variation smaller than a factor 100 in good agreement with experimental observations. This is why the dislocation density we used in the DD simulations varies with the grain size as reported in Table 2. In addition, as discussed by many authors (see for instance El-Awady, 2015), many experimental evidences show that it is unrealistic to expect that the dislocation density in the different grain sizes would be the same.

Since the interaction coefficient of the Taylor equation (see Section 3.1) varies with the initial dislocation density, the corresponding coefficient (usually called forest interaction strength or Taylor coefficient) is calculated from the simulations. The corresponding values are given in Table 2.

| Cubic grain size dimension $d$ (μm) | 1.25 | 2.5 | 5 | 10 |
|---|---|---|---|---|
| Dislocation loop size $l$ (μm) | 0.375 | 0.75 | 1.5 | 3 |
| Dislocation density $\rho_{ini}$ ($10^{12}$ m$^{-2}$) | 30 | 7.7 | 2 | 0.48 |
| Strain rate amplitude (s$^{-1}$) | 250 | 200 | 50 | 15 |
| Average dislocation strength $\alpha$ | 0.33 | 0.38 | 0.42 | 0.45 |

Table 2: Simulation conditions and initial dislocation microstructure definition as function of grain size. The Taylor coefficient is calculated from the simulated stress-strain curves at yield.

## 2.3  Simulated grain shapes

One of the main goals of this work is to quantitatively evaluate the impact of grain shape on the HP law. Hence, three different grain shapes are considered (see Figure 2). We consider cube, plate and needle grain shapes, to simulate the shapes frequently found in polycrystals. These shapes are close to those observed in recrystallized polycrystals with equiaxed grains, or lathes and needles observed during allotropic transformations (bainite, perlite and martensite, etc (Bhadeshia and Honeycombe, 2006)) and thin film (Yoshinaga et al., 2008).

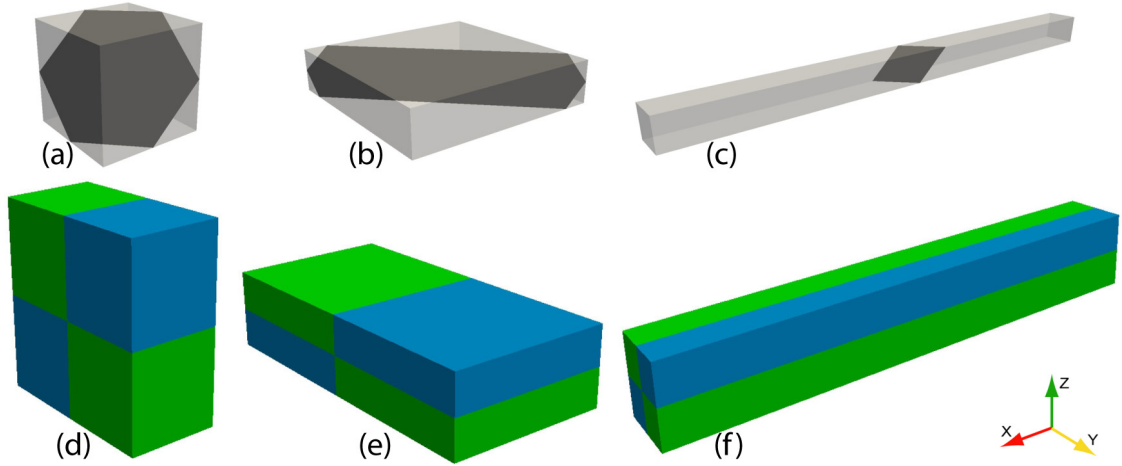

Figure 2: (a, b, c) the three grain shapes investigated with the simulations. One {111} FCC slip plane at the center of each grain is drawn to illustrate the variation of the slip plane area; (d, e, f) the three different types of *four-grained* simulated aggregates. Grains of the same color are mechanically loaded identically but contain different dislocation distributions.

The ratios between the linear dimensions of the three grain shapes are as follows: three almost equal dimensions (5;5;5) for the cube grain, two large dimensions (8;8;2) in the plate grain and one large dimension (31;2;2) in the needle grain. By convention, the grain size in all cases is defined as $d = \sqrt[3]{V}$.

## 3.  The grain size effects

Given the limitation of DD calculations, it is clear that DD cannot entirely address the grain size effect observed in polycrystals. However, a formulation similar to that in Eq. 1 represents a useful framework for the analysis and interpretation of our results. In this context, the friction stress $\sigma_o$ in the HP law accounts for the *intragranular* yield stress, as suggested by Cordero *et al.* (Cordero et al., 2016). $\sigma_o$ is material dependent and this quantity must be defined to calculate the HP term. We suppose in the simulations that $\sigma_o$ is equal to the yield stress of an ideal Taylor polycrystal, i.e. a polycrystal of random texture and very large grains. For such polycrystals, $\sigma_o$ is directly related to the Critical Resolved Shear Stress $\tau_c$ measured in a single crystal made with the same material. In addition, in the *one-grain* polycrystal simulations, $\sigma_o$ reduces to the yield stress of an infinitely large grain and the HP law becomes:

$$\sigma_y = \frac{\tau_c}{m} + Kd^n \qquad (2)$$

with $m$ is the highest Schmid factor of slip systems calculated in the simulated grain. Here, it must be noted that this is the main advantage of the *one-grain* simulations. Such simulations give the possibility to study the grain size effect at the slip system level and to reduce the HP law on the active slip systems to:

$$\tau_y = \tau_c + kd^n \qquad (3)$$

where $\tau_y = m\sigma_y$, $\tau_c = m\sigma_o$ and $k = mK$ the reduced HP constant.

To test the validity of Eq. 2 or Eq. 3, the plastic behavior of a single crystal is first determined to calculate the value of $\tau_c$ in all simulations. In single crystals, $\tau_c$ is controlled by forest interactions and is expected to follow the Taylor equation:

$$\tau_c = \tau_f + \alpha Gb\sqrt{\rho} \qquad (4)$$

where $\tau_f$ is a lattice friction (solid solution) accounting for dislocation interaction with impurities. In the present study, $\tau_f$ is fixed to the value of 2.5 MPa in the simulation to mimic common copper polycrystals with a low purity.

In the Taylor equation, the coefficient $\alpha$ is the forest strength coefficient, which was directly calculated by DD simulations (Madec et al., 2002). In the following, a simple procedure proposed by Devincre *et al.* (2006) is used to account for the variation of $\alpha$ with the dislocation density $\rho$:

$$\alpha = \frac{\ln\left(1/\alpha_{ref}b\sqrt{\rho}\right)}{\ln\left(1/\alpha_{ref}b\sqrt{\rho_{ref}}\right)}\alpha_{ref} \qquad (5)$$

where $\alpha_{ref}$ is evaluated to 0.4 at $\rho_{ref} = 10^{12}m^{-2}$ in the case of a uniform distribution of slip systems.

Since the initial dislocation density increases with decreasing grain size (see Table 2), the friction stress $\sigma_o$ in the HP law is following Eq.4 expected to increase with decreasing grain size. This variation in $\sigma_o$ cannot be neglected when calculating the HP term ($Kd^n$). This is why $\sigma_o$ must be assessed. In order to do so, simulations of copper single crystals with increasing initial dislocation density are reported in the next section.

### 3.1 Assessment of the single crystal behavior

For each tested initial dislocation density (see Table 2), three different tensile axes were considered: [001], [111] and [135]. As explained in section 2, single crystal simulations are made replacing in the *one-grain* simulations the impenetrable boundary (IB) condition by the transparent boundary (TB) condition.

As illustrated in Figure 3a, the resolved shear stress $\tau$ computed in the single crystal simulations does not evolve rapidly with the strain. Hence, strain hardening is low. To complete this general remark, it can be seen that the [111] orientation as a higher strain hardening rate than the [001] orientation while both are under multiple-slip condition. On the other hand, the orientation [135], leading to single slip condition, has non-visible strain hardening. The computed strain hardening rates are: $\theta_{001} \approx 170$MPa, $\theta_{111} \approx 320$MPa and $\theta_{135} \approx 0$. Such behavior is in good agreement with experiments where strain hardening is in the order of G/200 in multislip and G/3000 in single slip conditions (Devincre et al. 2008). In these conditions, it is possible to evaluate without ambiguity $\tau_c$ in each simulation. The computed values are plotted as a function of the initial dislocation density in Figure 3b. As expected, all the simulation results are in good agreement with the predictions of the Taylor equation (eq. 5), when the drift of the forest strength coefficient $\alpha$ associated to the variation of the dislocation line tension with the dislocation density $\rho$ is considered.

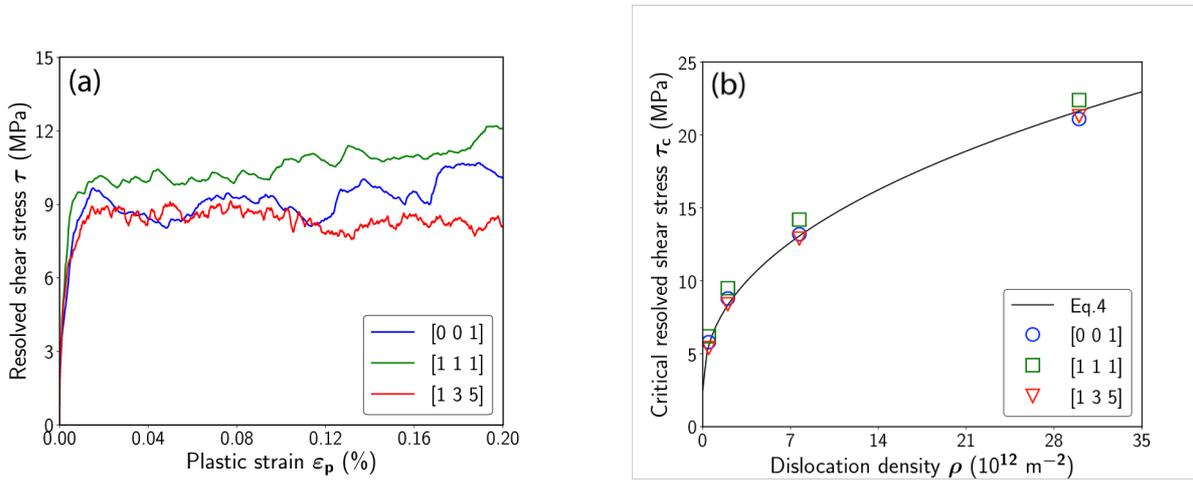

Figure 3: (a) Stress-strain curves of 3 single crystal simulations with 3 different crystal orientations. Computations are made using transparent boundaries of simulation volume of 5μm size. (b) Evolution of the CRSS as a function of the initial dislocation density (see Table 2). The full line is the prediction of the Taylor equation (Eq. 4).

Two additional observations can be made from Figure 3b: (i) $\tau_c$ is only slightly affected by the orientation of the tensile axis. Hence, the nature and number of active slip systems have little influence on the yield stress of single crystal. (ii) Predictions of Eq. 4 constitute a lower bound for the values of $\tau_c$ obtained in DD simulations. This is the signature of the effective stress needed to impose plastic strain by imposing a finite velocity of mobile dislocations, which necessarily causes a shift from the quasi-static critical stress prediction made with the Taylor equation. Nevertheless, the difference between the computed values and the model predictions never exceeds a few percent.

## 3.2 Impenetrable boundaries and plastic strain hardening

In a second step, comparison is made between the results of the *single crystal* and the *one-grain* polycrystal using the same simulation volumes and the same initial dislocation configurations. Going from one simulation to the other implies only to switch from TB to IB condition. In Figure 4 we show an example of the evolution of the applied stress (a) and the dislocation density (b) as a function of the plastic strain in a cubic volume of 5 μm loaded in the [001] orientation. One can easily see that the flow stress and the dislocation density increase with strain much faster in the *one-grain* simulation than in the *single crystal* simulations. This is a direct outcome of IB condition and a signature of the size effect. The large dislocation storage rate observed in the *one-grain* simulation is a consequence of dislocation accumulation at the GBs.

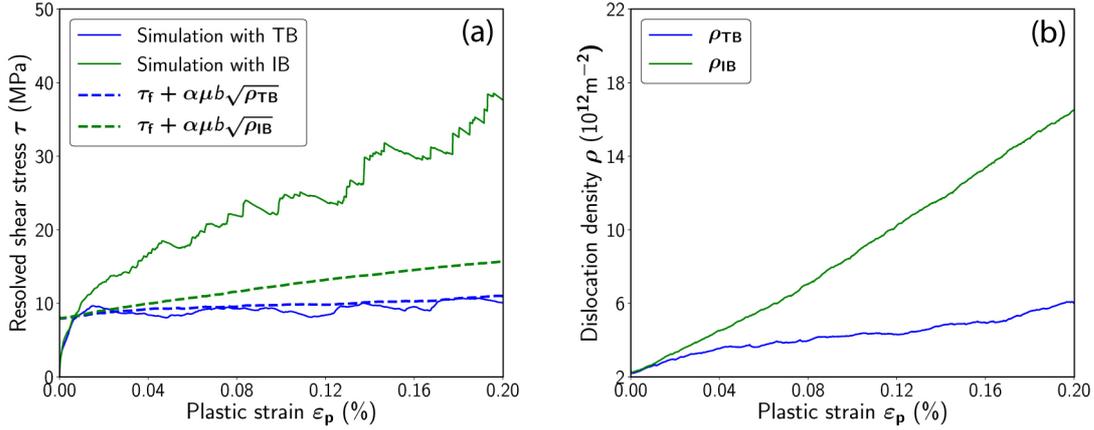

Figure 4. Comparison between simulations using transparent boundaries (TBs) and impenetrable boundaries (IBs) in a 5 μm cubic simulation volume loaded in the [001] tensile axis: (a) stress-strain curves and (b) dislocation density as a function of plastic strain. Dashed lines in (a) correspond to the prediction of Eq. 4 with the dislocation density recorded in (b).

Strain hardening calculated in the *one-grain* simulation is almost constant and amounts to approximately 13 GPa in Figure 4, which corresponds to approximately *G/3*. This hardening rate is 100 times larger than that found in the *single crystal* simulation. It is computed up to a plastic strain of 0.2%, i.e. the conventional elastic limit considered for the determination of yield stress in most experiments. As revealed in Figure 4a, the huge hardening rate we found in the *one-grain* simulation takes place from the very beginning of the plastic deformation and cannot be justified by an increase of forest hardening induced by the increase of the dislocation storage rate (see § Discussion).

### 3.3 Hall-Petch effects in the *one-grain* simulations

Now, we investigate the effect of the grain size on the yield stress using the *one-grain* simulations. Variations of the grain size are tested from 1.25 *μm* to 10 *μm* (see Table 2). In this section, the grains are cubic and tensile tests are made in different orientations: [001], [111] and [135]. Examples of stress-strain curves computed for the *one-grain* simulations deformed in the [135] direction are shown in Figure 5a. Results of these simulations are compared with the single crystal behavior obtained with the same dislocation configuration, i.e. an initial dislocation density of $2.10^{12}$ m$^{-2}$. One sees that the flow stress increases substantially with decreasing grain size. Again, it must be emphasized that this strengthening cannot be explained by the modification of initial dislocation density with the grain size. Moreover, when comparing the different curves in Figure 5a, it appears that strain hardening is weakly sensitive to grain size in those simulations.

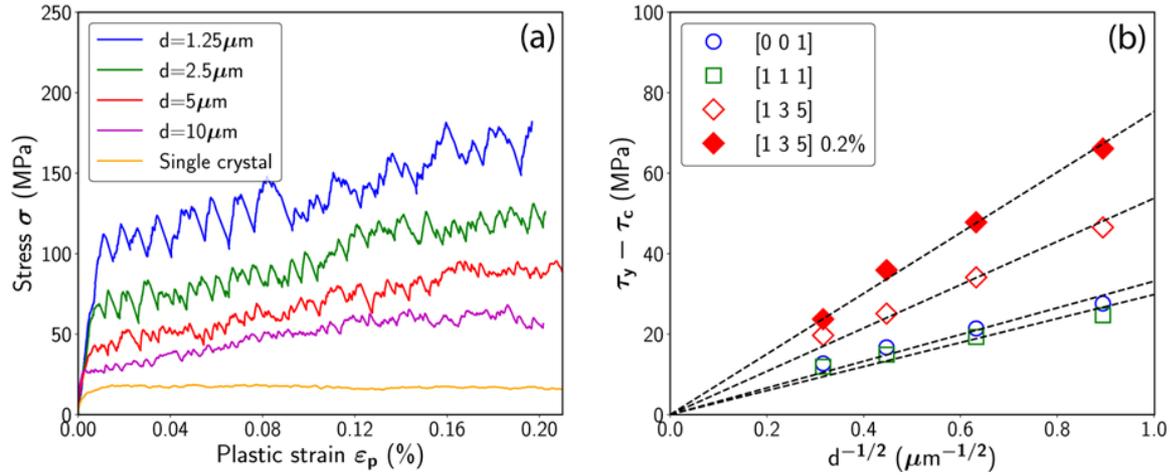

Figure 5 (a) Example of stress-strain curves obtained with the *one-grain* simulations for tensile tests along the [135] axis and for different grain sizes. The stress-strain curve simulated with TBs is added for comparison; (b) evolution of the HP term calculated at plastic strain equal to 0.1% and with the grain size of different orientations (full diamonds correspond to solutions obtained with 0.2% offset of plastic strain).

In order to investigate the grain size effect using Eq. 3, the critical stress $\tau_c$ computed from the single crystal simulation is subtracted from the resolved yield stress $\tau_y$ recorded in the *one-grain* simulations. Results of those calculations are plotted as a function of grain size $d$ in Figure 5b for the twelve simulations we performed. Simulation results systematically exhibit linear dependency on the inverse of the square root of the grain size. In addition, in the case of the *one-grain* simulations oriented for single slip deformation, the grain size effect is tested at two plastic strain amplitude (0.1% and 0.2%). From this last set of simulation results, it is clear that the slope of the linear fit increases with the selected plastic strain offset.

Hence, the HP constant $K$ given by the slope of the linear fit (see dashed lines in Figure 5b) varies with the loading axis and selected offset of the yield stress. Such variation is not a discrepancy of the simulations, but the consequence of the *one-grain* simulation condition. We will see in the discussions section that a realistic value of $K$ can be defined as an average of those obtained from the *one-grain* simulations with different plastic strain amplitudes and different grain orientations.

## 4. Results on the grain shape effects

### 4.1 The *one-grain* simulation

The same set of simulations described in the previous section was repeated with grains of plate and needle shapes. The increase in the critical stress as a function of the grain size for the three tensile directions is depicted in Figure 6. For the sake of clarity, simulation results are split into three figures depending on the loading axis.

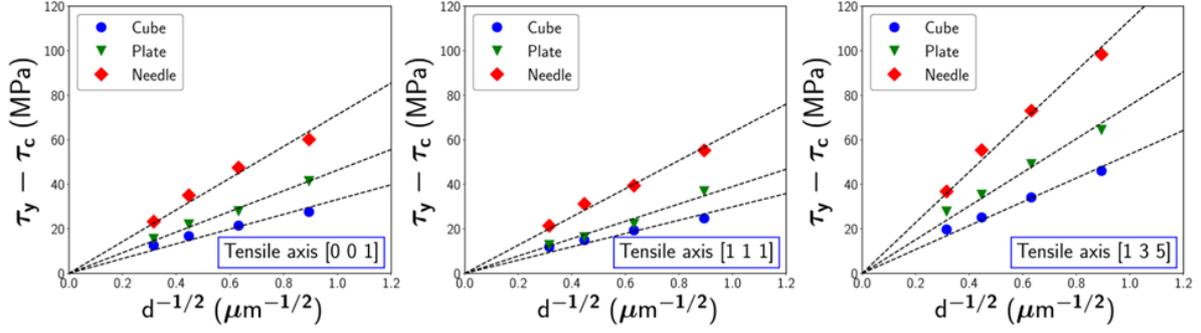

Figure 6. Variation of the HP term as function of the grain size for the three simulated grain shapes and loading axis.

As can be seen, the HP term, i.e. the term ($\tau_y$ - $\tau_c$), is always approximately proportional to $1/\sqrt{d}$. The HP law seems to be still valid in the case of non-equiaxed grains. However, the slope of the curves, i.e. the HP constant *K*, depends strongly on the grain shape. For the three investigated shapes, *K* has the largest value in the case of needles and weakest for cubic grains. On the other hand, the HP constant is higher in the single slip condition ([135] orientation), while it remains substantially the same in multislip conditions ([001] and [111] axes).

## 4.2 *Four-grain* simulations

In the previously reported simulations, the HP effect is investigated with particular aggregates made of one grain surrounded by its periodic images. We can suspect this configuration to induce artifacts due to the equal slip activity on both sides of the GBs. To investigate the effect of plastic strain incompatibility at GBs, the behavior of *four-grain* aggregates (described in Sec. 2) is now investigated. For the sake of simplicity, these simulations are performed at imposed stress rate (stress-controlled mode). As the previous simulations have been made at constant strain rate (strain controlled mode), we first check the consistency between the two controlling modes in the simulation. In Figure 7a, the results of *one-grain* simulations with a 2.5µm cubic grain deformed along the [001] and the [111] axis are shown as a function of loading modes. From such tests, we see that the two controlling modes give basically the same mechanical response and both controlling modes can be used indiscriminately in the aggregate simulations.

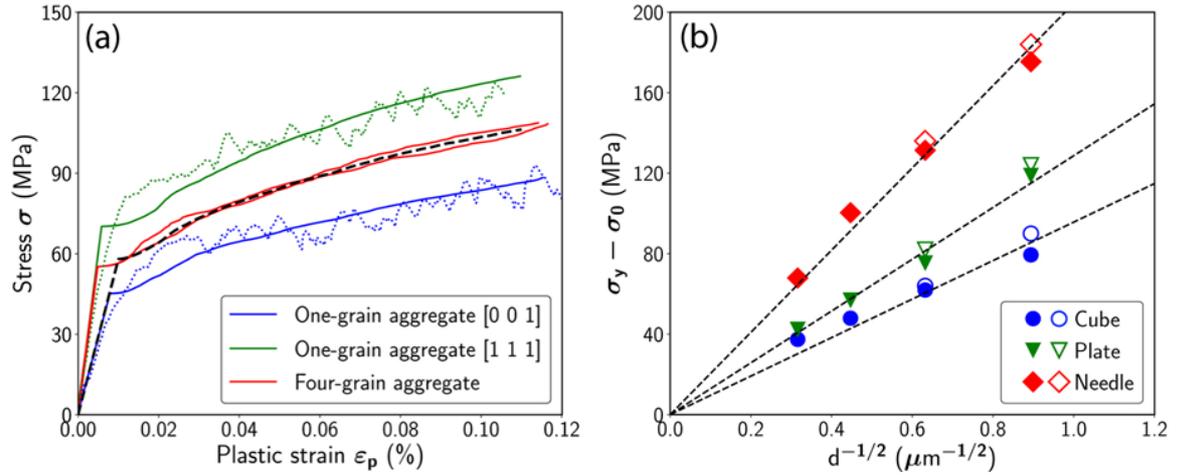

Figure 7. (a) Comparison between stress-controlled (full lines) and strain-controlled (dotted lines) modes for the *one-grain* simulations. The cubic grain is 2.5µm size and two loading axes ([100] and [111]) are considered. For comparison, two stress-strain curves calculated with the *four-grain* simulations using the same grain shape, same size and two loading orientations are reproduced in red curves. The two reported *four-grain* simulations only differ in the initial dislocation distributions. The black dashed line is the average of the *one-grain* simulations; (b) evolution of the HP term calculated with the *four-grain* simulations (open symbols) and calculated as an average of the *one-grain* simulations made with the [100] and [111] loading axis (full symbols).

In addition, in Figure 7a two stress-strain curves taken from simulations of *four-grain* aggregate with grains alternatively oriented in [001] and [111] and different in their random initial dislocation distribution are presented. Again, it can be seen that both stress-strain curves are very close to each other. This ensures that the initial dislocation microstructure has little influence on the simulation results.

Furthermore, it can be noted in Figure 7a that the *four-grain* aggregate stress-strain curves are close to the average of the stress-strain curves (black dashed line) of the *one-grain* simulations of the [111] and [001] orientations.

Results relative to the evolution of the HP term calculated with the *four-grain* simulations as a function of grain size and for different grain shapes (cube, plate and needle) are reported in Figure 7b. In this figure, we see that the linear dependency of the HP term on the inverse of the square root of the grain size applies in all cases. Also, the values of the HP constant $K$ (slope of the straight lines in Figure 7b) vary substantially with the grain shapes. The largest value of $K$ is obtained with the needle shapes and the lowest for the cubic grains.

### 4.3 The Bauschinger effect

Since most of the dislocation content in the grains are accumulated at grain boundaries, it is interesting to investigate a possible Bauschinger effect in our simulation conditions. To do so, a specific simulation is performed on the 5 μm grain using the impenetrable boundary condition. The tensile axis is [135] and the loading is reversed at 0.3% of plastic slip. The corresponding stress-strain curve is given in **Erreur ! Source du renvoi introuvable.**.

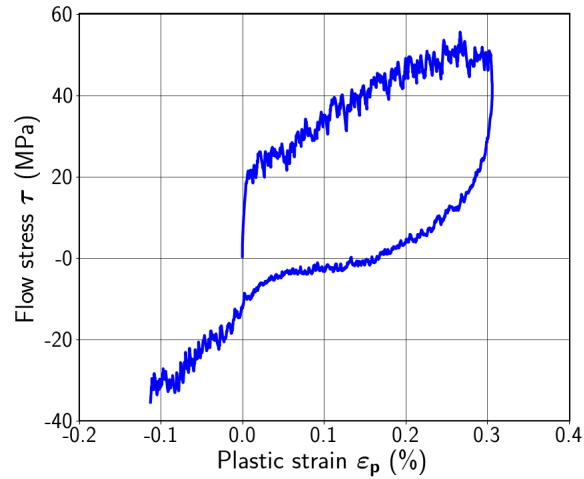

Figure 8: Stress strain curve associated to the Bauschinger test made for the 5 μm grain size simulation and the loading axis [135].

As can be noticed on the curve, the strain reversion leads first to a vertical drop of the stress with respect to the plastic strain axis. This elastic unload step is followed by approximately 0.1% of plastic strain reversal. This plastic strain relaxation reveals that one third of the dislocation microstructure accumulated at the grain boundary is pushed back when unloading. In addition, plastic strain in the compression direction is observed at stress amplitude smaller than -5MPa. The Bauschinger effect we calculate is thus very strong and indicates the presence of a strong back-stress with an amplitude close to the flow stress. This test clearly shows that kinematic hardening is predominant with respect to the isotropic hardening in our simulations.

## 5. Discussion

According to the usual interpretation of the HP effects (Armstrong et al., 1962; Cordero et al., 2016), the yield stress of polycrystals is composed of two linearly superposed contributions: the intragranular one, represented by the friction stress $\sigma_o$ and the intergranular component represented by the HP term ($Kd^n$). The main objective of this work is to investigate the evolution of the HP term as a function of the grain size and shape when assuming that GBs are impenetrable barriers to dislocation motion (i.e. in the relatively low stress and low strain conditions). To do so, the HP term must be isolated from the friction stress $\sigma_o$ in the simulated flow stress. In the periodic aggregates models we investigated (*one-grain* and *four-grain* simulations), $\sigma_o$ is by definition the flow stress of the same aggregates, but with very large grain size. Because of DD simulation limitations, it is not possible to compute directly $\sigma_o$. This is why the friction stress of each simulation condition must be evaluated. In the case of the *one-grain* simulations, $\sigma_o$ becomes the yield stress of a large single crystal oriented in the same direction as the grain in the simulation. The results reported in Figure 4 prove that $\sigma_o$ can be precisely evaluated by the Taylor equation (Eq. 4) provided that the variation of the forest strength $\alpha$ with the initial dislocation density is taken into account. In the *four-grain* simulations, the results reported in Figure 7 show that $\sigma_o$ can be considered as the average of the Taylor equation predictions of the critical stress of all grain orientations present in the aggregate.

## 5.1 Size effects in equiaxed grains

In all simulations performed in this work, we calculated the HP term and examined the dependency on *d*, as given in Eq. 2. Our results show good alignment with $d^n$ when -0.7 < *n* < 0.3. This is in partial agreement with the statistical analysis made by (Dunstan and Bushby, 2014) which shows that there is no conclusive experimental evidence of the linearity with n = -0.5 dependency. However, when we impose that the fitting lines pass by the origin, only values of -0.5 < *n* < -0.4 provide good fit to our results. This is why, we consider in this discussion the exponent to be equal to -0.5. The good alignment with $1/\sqrt{d}$ is evidenced, in Figure 5 and Figure 6 for the *one-grain* simulations and in Figure 7 for the *four-grain* simulations. This choice is consistent with the initial HP relation (Hall, 1951) and many experimental investigations (Armstrong et al., 1962; Cordero et al., 2016; Li et al., 2016).

Since dislocation storage strongly increases when grain size decreases, it is interesting to check whether the obtained size effects can be explained by an isotropic hardening resulting from the increase in dislocation density. To this end, we first calculate the forest strengthening as a function of the recorded dislocation density. Dislocation density increases basically only on active slip systems. Therefore, the use of the Taylor equation with an average forest strength α is not justified at finite strain. Instead, the tensor form of the Taylor equation accounting for dislocation density on each slip system proposed by Franciosi *et al.* (Franciosi et al., 1980) is preferable:

$$\tau^s = \tau_f + \mu b \sqrt{\sum_{i=1}^{12} a^{si} \rho^i}. \qquad (6)$$

In Eq. 6, $a^{si}$ is the interaction matrix known from previous DD simulations investigations (Hansen et al., 2010), accounting for the interaction strength between slip system *i* and *s*. Since dislocation density evolution on every slip system is recorded in the simulations, Eq. 6 gives a precise evaluation of the flow stress all along the stress strain curves. This is done in Figure 4a, where predictions of Eq. 6 are plotted in dashed lines for a single crystal simulation (transparent boundaries) and for the equivalent *one-grain* simulations (impenetrable boundaries). It can be seen that Eq. 6 correctly predict the flow stress of the single crystal simulation but not that of the aggregate simulation, where the predicted shear stress is far below the recorded flow stress. Indeed, in the presence of IBs a large number of dislocations (with the same Burgers vector and direction) is accumulated close to the boundaries. This phenomenon of polarized dislocation structure at GBs leads to a strong intergranular strengthening that cannot be predicted by the Taylor equation, used to calculate forest strengthening.

Another feature of interest is the effect of grain size on the evolution of the dislocation density storage. It is interesting to compare the dislocation storage computed within the DD simulations with a simple geometrical analysis. Neglecting forest interactions, every dislocation loop accommodates in cubic grains a shear increment $d\gamma \approx bd^2/V$, and contributes to an increase in dislocation density $d\rho \approx 4d/V$. The rate of dislocation storage is thus $d\rho/d\gamma \approx 4/(bd)$.

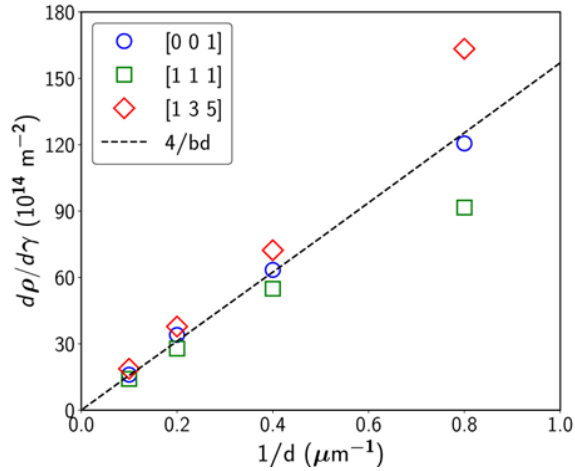

Figure 9: Rate of dislocation storage as a function of grain size. The dashed line represents the geometrical prediction proposed in this work.

This prediction is compared with simulation results in Figure 9, gathering the results of all the *one-grain* simulations with grains of cubic shape and different sizes and orientations. It can be seen that the storage rate recorded in DD simulations is inversely proportional to the grain size. Furthermore, the above geometrical prediction is in good agreement with DD results, which confirm that almost all stored dislocations are accumulating at the IBs. In other terms, we checked that, in the present simulation conditions, dislocation storage in the bulk of the grains is almost negligible with respect to that at GBs. The forest dislocation storage term ($\alpha\sqrt{\rho}/b$) frequently used in the literature when modeling the HP effect is not justified. These results are in full agreement with the classical Kocks-Mecking formulation of crystal plasticity simulations of polycrystal behavior by Haouala et al (Haouala et al., 2018).

## 5.2 Estimation of the HP constant *K*

As discussed before, the HP exponent *n* is found close to -0.5 in all simulations. Therefore, the square root dependency is adopted in the following discussion. By subtracting the flow stress of large single crystal ($\sigma_o$) estimated with Eq. 6, it is possible to calculate the HP constant *K* as a function of grain size and shape. For the *one-grain* simulations of cubic shape, the computed values of *K* (in MPa$\sqrt{m}$) is found to vary with respect to the tensile axis: 0.1 for the [135] axis, 0.06 for the [111] axis and 0.07 for the [001] axis. This variation is expected because the arrangement of dislocations accumulated at the IBs depends on the number of active slip systems.

As discussed in the introduction, the evolution of the flow stress of one grain or small grain clusters as a function of the grain size and shape is only one feature of the grain size effects in real polycrystals. Therefore, it is useful to compare the value of *K* calculated in the simulations with experimental observations. To do so, one must bear in mind that the experimental values are usually obtained on polycrystals with many grain orientations, while we have only one grain orientation in the *one-grain* periodic simulation. Using the Taylor hypothesis, which has been confirmed in copper (Kocks et al., 2000), the macroscopic stress becomes the average of the loading stress of the different constituent grains. Considering that the three simulated orientations are representative of common polycrystal texture, the HP constant measured in experiment can be compared with the average of the *K* values obtained in the simulations. The simple average of our results is *K* = 0.092 MPa m$^{-0.5}$, which is below the values reported in the literature (in MPa$\sqrt{m}$): 0.14 (Thompson et al., 1973), 0.19 (Meyers et al., 1995), 0.11 (Feltham and Meakin, 1957), 0.15 (Thompson and Backofen, 1971), 0.16 (Hansen and Ralph, 1982). The latter values yield an average experimental value *K* = 0.15 MPa$\sqrt{m}$, which is 39% larger than that estimated with the simulations.

Additional thinking suggests that the difference observed between simulations and experiments may be due to the selection of the 0.1% strain offset for the definition of yield in the simulations, while it is 0.2% strain in the experimental investigations. The 0.1% offset was considered in order to reduce the significant computation times. Examining the stress-strain curves of the *one-grain* simulations (see for example Figure 5a), it is clear that the flow stress increases strongly and monotonously with the strain. The value of *K* must thus increase when fixing the offset of yield strain at 0.2% in the simulation. Only four simulations with grains oriented parallel to the [135] tensile axis was conducted to 0.2% of plastic strain. The HP terms calculated for these simulations are reported in Figure 5b (full diamond marks). The corresponding HP constant *K* is 0.16 MPa$\sqrt{m}$, which better matches with the experimental values. It is worth noting that the presence of solute element in the grain boundary layer increases the stability of grains, which is known to raise the value of *K* in industrial Cu alloys compared to pure Cu (see for ex. Thompson et al. 1973). This finding has recently been confirmed in atomistic simulations of the Cu-Al systems by Borovikov et al. (Borovikov et al. 2017).

**5.3  Effect of grain shape**

As revealed in Figure 6, for all grain orientations the HP constant *K* increases strongly when the grain shape considered in simulation is not cubic. The largest value of *K* is found with the needle shape, with a value two times larger than that obtained for cubic grains. This indicates that the HP effect increases when the aspect ratio of the grain is different from 1. We propose the following model to account for this property.

Let us consider grains in the form of parallelepipeds with three independent dimensions. As the accumulation of polarized dislocations close to IBs and its associated back stress is at the heart of the interpretations of the HP effect, we consider a dislocation loop expanding on a slip plane from the center of the grain. Its final length *C* is equal to the circumference of the surface of the slip plan bounded by the grain faces. During expansion, the accommodated plastic deformation is proportional to the area *S* of this surface. If the back stress is mainly associated with the polarized dislocation accumulated against IBs, it must be proportional to *C*. On the other hand, during deformation, the back stress increase with deformation is likely to be inversely proportional to *S*. Consequently, we expect *K* to be proportional to *C/S*. This factor has dimensionality [m$^{-1}$] and has to be normalized by the corresponding factor of a cubic grain shape to recover a factor of unity in the case of equiaxed grains. One can thus define a shape factor as:

$$\psi = \frac{C}{S} \frac{S_{cube}}{C_{cube}} \qquad (7)$$

When considering FCC slip system symmetry and parallelepipedic grains with boundaries perpendicular to <100> directions, this shape factor takes the value ψ = 1 (cube), ψ = 1.51 (plate) and ψ = 2.52 (needle). The HP terms deduced from Figure 6 are plotted in Figure 10 as a function of $1/\sqrt{d}$ weighted by the shape factor. It can be seen that all data now collapse on the same straight line going through zero at very large grains. This confirms that the effect of grain shape revealed in the simulations is well taken into account using the shape factor defined in Eq. 7.

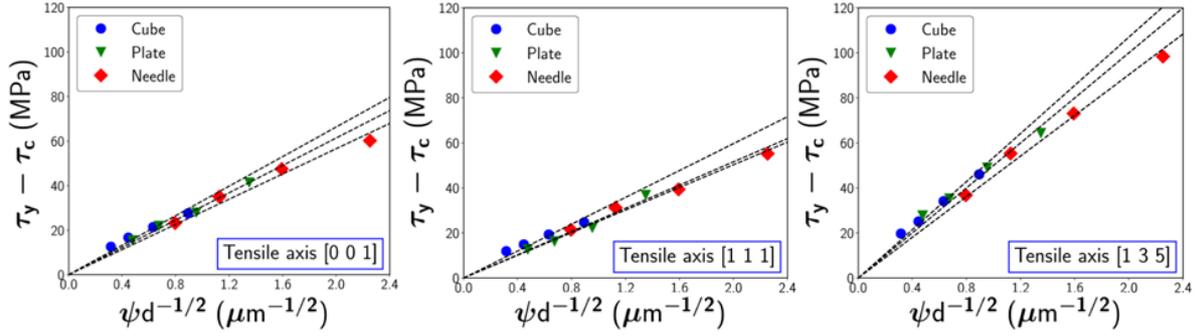

Figure 10. Evolution of the HP term with the grain size multiplied by the shape factor ψ for the three tested grain orientations and three grain shapes.

In other words, our simulations suggest that in order to apply the HP relation in the case of non-isotropic grain shapes of a volume *V*, one should define an effective grain size $d_{eff}$ accounting for the shape factor given in Eq. 7. The present model is based on the calculation of the average dislocation loop length *C* and the average area swept by dislocations *S*, suggesting the following generalized Hall-Petch relation:

$$\sigma_y = \sigma_0 + \frac{K}{\sqrt{d_{eff}}} \qquad \text{with} \qquad d_{eff} = \left(\frac{4S}{C}\right)^2 \times \frac{1}{\sqrt[3]{V}} \qquad (8)$$

Eq. 8 involves three independent parameters because it is developed for parallelepipedic grains of three independent dimensions. Of course, for the geometries used in our simulations (needles and plates), Eq. 8 reduces to only two independent parameters.

### 5.4 Plastic strain incompatibility in the *four-grain* simulations

As shown in Figure 7, the stress-strain curve of *four-grain* simulations including [001] and [111] grains is close to the average of the stress-strain curves calculated with the *one-grain* simulations for the [001] and [111] grain orientations. This result is surprising, since every grain is surrounded by grains of different slip activity in the *four-grain* simulations, while slip activity is identical on both sides of the boundaries in the *one-grain* simulations. The long range stress field associated with the polarized dislocation density accumulated at IBs is then expected to be different. Thus, in the case of periodic cluster simulations, it seems that the flow stress is little sensitive to the details of the grain neighborhood; the overall behavior is a simple average of the individual responses of its grains. This result supports the assumption made in homogenization methods (Roters et al., 2010) and XRD observations made on grain rotations in polycrystals deformation (Poulsen et al., 2003; Winther et al., 2004), at least, at the beginning of plastic deformation.

Consequently, it is relevant to check if the shape factor defined in the previous section still applies to the *four-grain* simulations with non-cubic grains (see results in Figure 7b). The corresponding HP analysis as a function of the effective grain size $d_{eff}$ is plotted in Figure 11a in which $d_{eff}$ was normalized by the norm of the Burgers vector $b$.

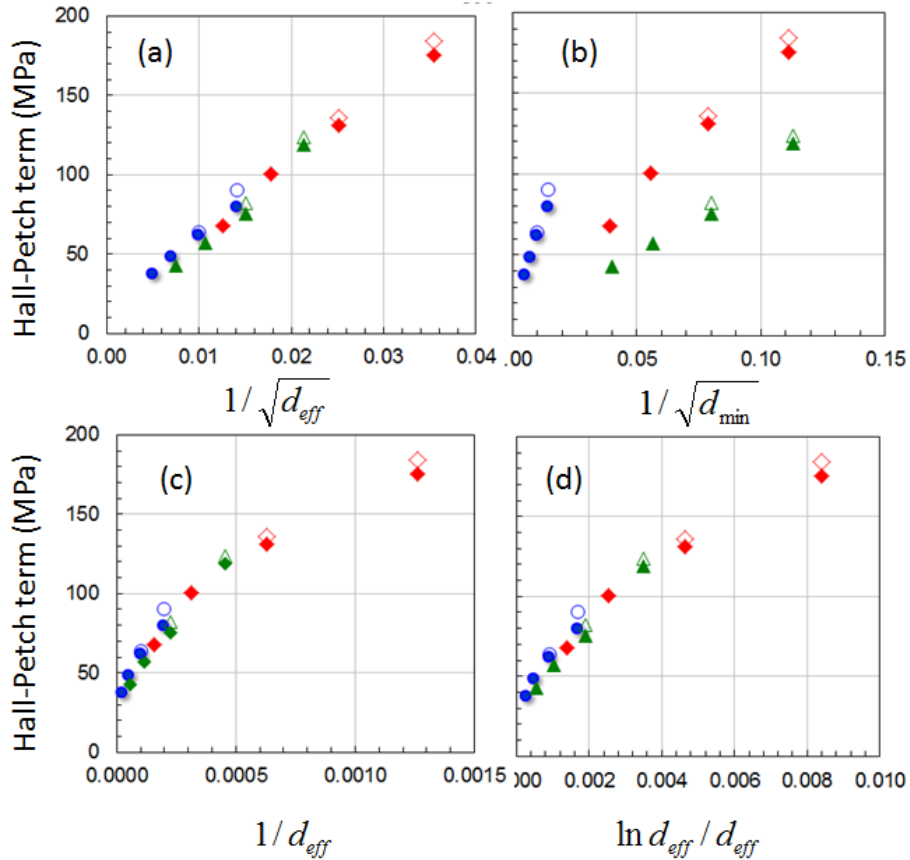

Figure 11. HP term vs different functions of the effective grain size $d_{eff}$ and the smallest edge length of grain $d_{min}$, including simulations results on cubic (circles), plate (triangles) and needle (diamonds) shapes on one-grain (full symbols) and *four-grain clusters* with [001] and [111] orientations (open symbols).

It can be seen that results of the *four-grain* simulations with different grain shapes (cube, plate and needles) are again aligned on the same line passing roughly by the origin. The corresponding HP constant $K$ is equal to 0.087 MPa$\sqrt{m}$, which is quite close to the average of the values obtained for the [001] and [111] *one-grain* simulations (given in full symbols in Figure 11). In addition, this value is lower than that obtained from averaging the $K$ values we obtained in section 5.2 when accounting for all the *one-grain* simulations. This discrepancy comes from the absence of grains oriented for single slip condition (e.g. <135> orientation) in the *four-grain* simulations. Indeed, we show in Figure 6 that these grains induce the largest HP effect. Hence, our simulations show that polycrystal aggregates with texture endorsing single slip deformation in the grains are expected to provide enhanced HP effect at the yield strain.

Following the analyses made by Li *et al.* (Li et al., 2016), where it has been shown that experimental results fit well on different functions : $1/\sqrt{d_{eff}}$, $1/d_{eff}$ and $\ln d_{eff}/d_{eff}$, it is important to check if this feature applies also on our results. In Figure 11c and d, we plot the values of the HP term given in Figure 11a as a function of $1/d_{eff}$ and $\ln d_{eff}/d_{eff}$, respectively. Also, as sometime proposed in the literature, we plot in Figure 11b calculated values of the HP term as a function of $d_{min}$, the smallest edge length of grains. It can be noticed that any of these tests gives data aligned on a straight line passing by the origin. The trends observed in Figure 11c and d cut the y-axis at a significantly high value ($\approx$ 30 MPa). This feature cannot be easily rationalized using physical arguments. Consequently, the best rationalization of our results seems to be given using the inverse of the square root of the effective grain size.

## 6. Conclusion

In this paper, we investigated the flow stress of periodic grains and 4-grain clusters of different sizes and shapes with impenetrable boundaries. The following conclusions can be drawn from the DD simulations reported in this paper:

- The critical resolved shear stress of single crystals is correctly predicted by the forest model given by the Taylor equation providing that the logarithmic dependency on the dislocations density of the forest strength ($\alpha$) is appropriately accounted for.
- Strengthening induced by dislocations accumulated at GBs cannot by predicted using the Taylor equation. An alternative constitutive equation will be proposed in a forthcoming paper.
- The yield stress at 0.1% and 0.2% offset of plastic strain in grains of orientations <100>, <111> and <135> and shapes cube, plate and needle, is found to vary almost linearly with the inverse of the square root of the grain size.
- The grain size effect on the flow stress varies strongly with the grain orientation. The average of the HP constant values (*K*) calculated at 0.2% offset of plastic strain is close to the values reported in experiments on copper polycrystals. This suggests that the strong assumption we made when considering GBs as impenetrable captures the basic mechanical features that take place at low plastic strain and low stress.
- The HP constant varies strongly with the grain shape. A shape factor is proposed to account for this sensitivity and an effective grain size is defined in order to keep using the HP relation in the case of non-equiaxed grains.
- The mechanical response of periodic four-grain aggregates is well approximated by the average of responses of its constituent periodic grains.

Modeling the grain boundary interactions with moving dislocations (absorption, repulsion, emission) requires specific physically based constitutive rules for the boundary response. Such rules can only be addressed in atomistic simulations, where the atomic structures and thermodynamics of the boundaries are fully accounted for. On the other side, only DD simulation gives the possibility to reproduce the kinematic conditions associated to realistic dislocation patterns accumulated in the vicinity of grain boundaries. This information is essential to investigate the long-range internal stresses and stress concentrations affecting plastic deformation of polycrystals. Multiscale modeling of the inter- and intragranular features of plastic deformation of polycrystals is an exciting perspective that remains to be explored.

**Acknowledgment**

This work is partially funded from the Euratom research and training program 2014-2018 under grant agreement No 661913 (SOTERIA). This work also contributes to the Joint Program on Nuclear Materials (JPNM) of the European Energy Research Alliance (EERA).